\documentstyle[aps]{revtex}
\begin{document}

\begin{flushright}
{\large\bf $<<$Nuclear Physics B 546 [FS] (1999) 691-710$>>$}
\end{flushright}

\centerline{\Large {\bf Two Magnetic Impurities with Arbitrary Spins }} 
\centerline{\Large {\bf in Open Boundary $t-J$ Model }}\centerline{\bf 
Zhan-Ning Hu\footnote{{\bf E-mail: huzn@aphy.iphy.ac.cn}} and Fu-Cho 
Pu 
\footnote{{\bf E-mail: pufc@aphy.iphy.ac.cn}}} \centerline{Institute 
of
Physics and Center for Condensed Matter Physics,} \centerline{Chinese
Academy of Sciences, Beijing 100080, China}

\begin{center}
\begin{minipage}{5in}
\centerline{\large\bf   Abstract}
From the open boundary $t-J$ model, an impurity model is constructed 
in which magnetic impurities of arbitrary spins are coupled to the 
edges of the strongly correlated electron system. The boundary $R$ 
matrices are given explicitly. The interaction parameters between 
magnetic impurities and electrons are related to the potentials of 
the impurities to preserve the integrability of the system. The 
Hamiltonian of the impurity model is diagonalized exactly. The 
integral equations of the ground state are derived and the ground 
state properties are discussed in details. We discuss also the 
string solutions of the Bethe ansatz equations, which describe 
the bound states of the charges and spins. By minimizing the 
thermodynamic potential we get the thermodynamic Bethe ansatz 
equations. The finite size correction of the free energy 
contributed by the magnetic impurities is obtained explicitly. 
The properties of the system at some special limits  are  
discussed and the boundary bound states are obtained.



\smallskip

\end{minipage}
\end{center}

\newpage

\section{Introduction}

Strongly correlated systems have been of great interest over the past 
decade
because of their importance as the most fundamental systems related to 
the
theory of high- $T_c$ superconductors\cite{1200,1201} and these 
electron
systems pose many theoretical challenges due to their essentially
nonperturbative nature. The Hubbard chain and $t-J$ model are 
frequently
invoked as the basic models of the one dimensional strongly electron
systems. When the repulsion energy $U$ for two electrons located on the 
same
atom is much larger than the bandwidth of the electrons, the Hubbard 
model
reduces to the $t-J$ model with an occupation of one electron per site. 
Then
there are three states per site in the $t-J$ model. Also the low-energy
excitations of the three-band Hubbard model can be changed into an 
effective
one-band $t-J$ model\cite{91201,91202,91203}. The $t-J$ Hamiltonian is 
one
of the most attractive model to explain high temperature 
superconductivity
without any phonon pairing mechanism\cite{11101,11102}.

Recently great progress has been made for the one dimensional 
systems\cite
{review}, in particular for the quantum impurity problems such as Kondo
problem and tunneling in quantum wires. The theory of the Kondo 
effect\cite
{kkkk} was developed in the early 1960s to explain the puzzle of the
resistance of some metals, that is, the resistance start to increase 
as the
metal is cooled below a certain temperature. This is due to that the 
local
moments on the impurity atoms having an antiferromagnetic coupling to 
the
spins of the conduction electrons and the coupling becomes stronger as 
the
temperature falls. Now we know the impurities play an important role 
in the
strongly correlated electron compounds. Even a small amount of defects 
may
change the properties of the electron systems.

Exacts results, even for a simplified model exploring just some features 
of
the system, are always useful and provide a testing ground for 
approaches
intended for the full problem of higher order of complexity. Therefore 
it is
of interest to construct integrable systems of the strongly correlated
electron systems including impurities. Indeed, there has been a long
successful history to study the effects of the impurities in the 
many-body
quantum systems within the framework of the integrable systems\cite
{13801,13802,13803,13804,re05}, although the impurity usually 
destroys the
integrability of the model when it is introduced to an exactly solved 
model.
Andrei and Johannesson incorporated a magnetic impurity of the 
arbitrary
spin into the isotropic spin-$\frac 12$ Heisenberg chain with 
integrability
preserved. The results were extended to the Babujian-Takhtajan spin 
chain in
Refs. \cite{323301}. Bed\"{u}fig, E$\beta ler$ and Frahm\cite{bef2} 
solved
the integrable model with the impurity coupled with periodic $t-J$ 
chain\cite
{last01,last02,last03}, in which the impurity is introduced through a 
local
vertex. Schlottmann and Zvyagin introduced the impurity into 
supersymmetric $
t-J$ model\cite{foe01} via its scattering matrix with the itinerant 
electrons
\cite{schlo01,schlo02} . The Hamiltonian of the system and other 
conserved
currents can be constructed in principle by the transfer matrix. Zvyagin 
get
that the low field impurity behavior for the periodic correlated 
electron
chain coincides with that in the open chain up to mesoscopic 
corrections\cite
{prlzv}. They also discussed the magnetic impurities embedded in the 
Hubbard
model \cite{021sch} and a finite concentration of magnetic impurities
embedded in one-dimensional lattice via scattering 
matrices\cite{verry}. The
quantum impurity problem was also discussed by the use of the 
bosonization
technique \cite{liuye01,liuye02}.

The study of completely integrable quantum spin open chains\cite
{115501,115502,115503,115504,115505} is also an interesting subject. 
It
deals with the system in a finite interval. The pioneering works of
Cherednik and Sklyanin\cite{chere,115501} are the starting point for 
the one
dimensional exactly solved models where there is a variant of the usual
quantum inverse scattering method. The so-called reflection equations
appeared as a new ingredient to the Yang-Baxter equations for describing 
the
systems on a finite interval with independent boundary conditions on 
each
end. Using this method, a lot of integrable models have arisen\cite
{345601,345602,345603} such as the Hofstadter problem and the reaction
diffusion equations. The quantum-group-invariant transfer matrices and 
the
Hamiltonians can be given by the certain limits of the open-boundary
conditions\cite{893401,893402,893403,345601,115504}. Some models 
have been
solved with non-trivial generalizations of the methods developed by 
periodic
and twisted boundary conditions.

In 1993, Foerster and Karowski\cite{FoeKar} proposed the $spl_q(2,1)-$
invariant $t-J$ model with the quantum-group-invariant open boundary
conditions and it was generalized to the $SU_q\left( n\right) $-
invariant
chains by de Vega and Gonz$\stackrel{\text{'}}{\text{a}}$lez-Ruiz in 
Ref. 
\cite{15015}. Gonz$\stackrel{\text{'}}{\text{a}}$lez-Ruiz discussed 
also the
open boundary supersymmetric $t-J$ model commuting with the number 
operator $
n$ and $S^z$\cite{bpB468}. Recently, Zhou and Batchelor\cite{zhou} 
studied
the open boundary $t-J$ model via an analytic treatment of the Bethe 
ansatz
equations. Fan, Hou and Shi\cite{print0} formulated the eigenvalues and
eigenvectors of the $t-J$ model with reflecting boundary conditions by 
the
graded quantum inverse scattering method. The zero-temperature 
boundary
effects in an open $t-J$ chain with boundary fields were discussed by 
Essler 
\cite{essle}.

Now we know that the quantum inverse scattering method provides us also 
a
very useful technique to construct the impurity model with the 
preserving of
the integrability and this scheme has been adapted in Ref. \cite
{bef2,schlo01,021sch}. Very recently, the impurity model related to the 
$t-J$
model was studied also in Ref. \cite{foe03}. But, when the impurities 
are
embedded in the system with periodic boundary conditions, some 
unphysical
terms must be present in the Hamiltonian to keep the integrability 
though
they may be irrelevant. Notice the fact that the impurities cut the
one-dimensional system into `pieces' when they are introduced and (then) 
the
open boundary systems are formed with the impurities at the ends of the
systems. In fact, Gonz$\stackrel{\text{'}}{\text{a}}$lez-
Ruiz\cite{bpB468}
has pointed out that the boundary terms could take into account 
impurities
or magnetic fields located at the boundaries in 1994. So, the integrable
impurity models\cite{many05,many06,many07,many08,many09,many010} may 
be
studied with the use of open boundary conditions.

In this paper, we construct a Hamiltonian of the impurity model within 
the
framework of the open boundary $t-J$ chain. The magnetic impurities are
coupled to the edges of the strongly correlated electron system and have
arbitrary spins. The boundary $R$ matrices are given explicitly and 
satisfy
the reflection equation. They are compatible with the Yang-Baxter 
equation
in the bulk. These guarantee the integrability of our model including 
the
impurities with the arbitrary spins. The interaction parameters of the
magnetic impurities with the electrons are parameterized by the 
constants
which relate to the potentials of the impurities in order to contain 
the
integrability of the system. The Hamiltonian with the impurities is
diagonalized exactly by the use of the Bethe ansatz method and the Bethe
ansatz equations are obtained explicitly via the quantum inverse 
scattering
method. The integral equations of the ground state are derived and the
properties of the ground state are discussed in detail. The finite size
corrections of the ground state energy due to the impurities are 
obtained.
The distribution functions of the rapidities are given. The string 
solutions
of the Bethe ansatz equations are discussed and the coupled integral
equations of the excited state are obtained. By minimizing the
thermodynamical potential of the system with impurities, we get the
thermodynamical Bethe ansatz equations. The finite size correction of 
the
free energy is obtained for the two magnetic impurities. We study also 
the
properties of the system in the high-temperature and the low temperature
limits. The boundary bound states are constructed in our model.

The paper is divided into eight sections. In section 2 we give the
Hamiltonian of our impurity model where the generalized ``permutation
operator'' is presented. In section 3 the integrability conditions are
discussed and the boundary $R$ matrices are obtained. Our Hamiltonian 
is
diagonalized exactly and the Bethe ansatz equations of the impurity 
model
are obtained in section 4. In section 5 the coupled integral equations 
of
the ground state are derived and the ground state properties of the 
system
are discussed. In section 6, the string solutions of the Bethe ansatz
equations are discussed and the integral equations of the excited states 
are
obtained. We get also the thermodynamic Bethe ansatz equations. The 
finite
size correction of the free energy contributed by the two magnetic
impurities is obtained. In section 7, the properties of the system at 
the
high-temperature and the low temperature limits are discussed and the
boundary bound states are constructed. Conclusions follow in section 
8. In
order to complete the four integrable cases of the $t-J$ model, some
boundary $R$ matrices and the Bethe ansatz equations are given in 
Appendix A
and B, respectively.

\section{The Hamiltonian of the Impurity Model}

As is well known, the $t-J$ model describes electrons with spin $\frac 
12$
on an one-dimensional lattice and permits the nearest-neighbor hopping 
($t$)
of the electrons. The large on-site Coulomb repulsion makes that the
double-occupancy of every site impossible. There are two types of
interactions between the electrons on the nearest neighbor sites. One 
is the
spin exchange interaction ($J$) and the other is the charge interaction 
($V$
) independent of spin. Our starting point is the following Hamiltonian 
\begin{eqnarray}
H &=&-t\sum_{j=1}^{G-1}\sum_{\sigma =\uparrow \downarrow }(C_{j\sigma
}^{+}C_{j+1\sigma }+C_{j+1\sigma }^{+}\ 
C_{j\sigma })+J\sum_{j=1}^{G-1}{\bf S
}_j\cdot {\bf S}_{j+1}+V\sum_{j=1}^{G-1}n_jn_{j+1}  \nonumber \\
&&+J_a{\bf S}_1\cdot {\bf d}_a+V_an_1+J_b{\bf S}_G\cdot {\bf 
d}_b+V_bn_G,
\label{e1}
\end{eqnarray}
where $C_{j\sigma }^{+}(C_{j\sigma })$ is the creation (annihilation)
operator of a conduction electron with spin $\sigma $ on the site $j$; 
${\bf 
S}_j=\frac 12\sum_{\sigma ,\sigma ^{\prime }}C_{j\sigma }^{+}\sigma 
_{\sigma
,\sigma ^{\prime }}C_{j\sigma ^{\prime }}$ is the spin operator of the
conduction electron; $d_{a,b}$ are the spin operators of the impurities; 
$
J_{a,b},V_{a,b}$ are the interaction constants of the impurities with 
the
electrons and the scattering potentials of the magnetic impurities ,
respectively; $n_j=C_{j\uparrow }^{+}C_{j\uparrow }+C_{j\downarrow
}^{+}C_{j\downarrow }$is the number operator of the conduction 
electrons; $G$
is the length ( or site number ) of the system. For the bulk of the system,
the scattering matrices $S$ of the electrons satisfy the Yang-Baxter
equation: 
\begin{equation}
S_{12}(k_1,k_2)S_{13}(k_1,k_3)S_{23}(k_2,k_3)=S_{23}(k_2,k_3)S_{13}
(k_1,k_3)S_{12}(k_1,k_2)
\label{facto}
\end{equation}
which gives the factorization condition for the integrability of the 
free
boundary and periodic boundary $t-J$ models. For the open boundary 
system,
the reflection equation\cite{chere,115501} should also be satisfied, 
i.e. 
\begin{equation}
S_{12}(k_1,k_2)\stackrel{1}{R}(k_1)S_{12}(k_1,-
k_2)\stackrel{2}{R}(k_2)=
\stackrel{2}{R}(k_2)S_{12}(k_1,-
k_2)\stackrel{1}{R}(k_1)S_{12}(k_1,k_2),
\label{byb}
\end{equation}
where the boundary $R$ matrices are written down as 
\[
\stackrel{1}{R}(k_1)=R(k_1)\otimes id_{V_2},\,\qquad \stackrel{2}{R}
(k_2)=id_{V_1}\otimes R(k_2) 
\]
for matrix $R\in End(V).$

To consider magnetic impurities of arbitrary spins, we define an 
operator $
P_d$ as 
\begin{equation}
P_d=\frac 1{\sqrt{1+4l\left( l+1\right) }}\left( 1+4S\cdot d\right)
\end{equation}
where $S$ and $d$ are the spin operators of the electron and impurity,
respectively. It has the property $P_d^2=1$ and degenerates to the 
ordinary
permutation operator when the impurity spin $d$ takes its value $l=1/2.$ 
We
call $P_d$ the generalized ``permutation operator''. The Hamiltonian 
may be
written in the form

\begin{equation}
H=-\sum_{j=1}^N\left( T_j^{+}+T_j^{-}\right) +\sum_{j=1}^N\left(
K_j+K_{aj}\delta _{x_j,1}+K_{bj}\delta _{x_j,G}\right)
\end{equation}
where we have set $t=1.$ The translation operators $T_j^{\pm }$ are 
defined
as\cite{schultz01} 
\[
T_j^{\pm }\Psi (x_1,\cdots ,x_j,\cdots ,x_N)=\Psi (x_1,\cdots ,x_j\pm
1,\cdots ,x_N) 
\]
where $\Psi (x_1,\cdots ,x_j,\cdots ,x_N)$ is the wave function of $N$
conduction electrons. The operator $K_j$ acts on the wave function $\Psi
(x_1,\cdots ,x_j,\cdots ,x_N)$ in the form 
\[
K_j\Psi (x_1,x_2,\cdots ,x_N)=\sum_{i=1}^N\delta 
_{x_j,x_i+1}K_{ij}\Psi
(x_1,x_2,\cdots ,x_N) 
\]
where $K_{ij}=V-\frac J4+\frac J2P_{ij}$ denotes the interactions 
between
the conduction electrons. The permutation operator $P_{ij}$ permutes 
the
spins of the $i$ -th and $j$-th electrons. The interactions between the
impurities with arbitrary spins and the electrons are given by the 
relations 
\[
K_{aj}=V_a-
\frac{J_a}4+\frac{J_a\sqrt{1+4s_a\left( s_a+1\right) }}4P_{aj}, 
\]
\[
K_{bj}=V_b-
\frac{J_b}4+\frac{J_b\sqrt{1+4s_b\left( s_b+1\right) }}4P_{bj} 
\]
with 
\begin{equation}
P_{aj}=\frac 1{\sqrt{1+4s_a\left( s_a+1\right) }}\left( 1+4S_j\cdot
d_a\right) ,  \label{p001}
\end{equation}
\begin{equation}
P_{bj}=\frac 1{\sqrt{1+4s_b\left( s_b+1\right) }}\left( 1+4S_j\cdot
d_b\right)  \label{p002}
\end{equation}
where ${\bf S}_j$ are the spin operators of the conduction electrons; 
$d_a$
and $d_b$ are the spin operators of the magnetic impurities with the
arbitrary spin values $s_a$ and $s_{b,}$ respectively. The boundary $R$
matrices of our impurity model will be given explicitly in the following
section.

\section{The Boundary R Matrix and Integrability Conditions}

The boundary $R$ matrices satisfy the reflecting Yang-Baxter equation 
for
the open boundary system and the reflection equation is compatible with 
the
factorization condition. These give the integrability of the one-
dimensional
many body system. In our model, the boundary $R$ matrix should take the 
form 
\begin{equation}
R=\exp (i\varphi )\frac{q-ic-i\left( 2l+1\right) P_d/2}{q+ic+i\left(
2l+1\right) P_d/2},
\end{equation}
where $P_d\;$is the generalized ``permutation operator'', $q=\frac 
12\tan 
\frac k2$ and $c$ is an arbitrary constant. Here we have set $J=2$ and 
$
V=3/2.$ The other integrable cases will be given in the Appendix. Put 
$
K_{a(b),j}$ $=$ $m+nP_d$ . From the boundary Yang-Baxter relation we 
have
that 
\[
2n\left( \frac 2{2l+1}q^2+\frac 2{2l+1}c^2-\frac{2l+1}2\right) \tan 
\frac k2
\quad \quad \qquad \qquad 
\]
\begin{equation}
+q\left[ \left( m-1\right) ^2-n^2\right] \tan ^2\frac 
k2+q\left[ \left(
m+1\right) ^2-n^2\right] =0.
\end{equation}
This relation gives the restricting condition imposed on the coupling
constants of our impurity model. It should be satisfied to maintain the
integrability of the system when the magnetic impurities with the 
arbitrary
spins are introduced to the $t-J$ model by coupling the impurities to 
the
edges of the system. From the above relation, we have that 
\[
m=\frac{\left( 2l+1\right) ^2+1-4c^2}{4\left( l\pm c\right) 
\left( l\mp
c+1\right) }, 
\]
\[
n=\frac{2l+1}{2\left( l\pm c\right) \left( l\mp c+1\right) }. 
\]
The phase factor in the boundary $R$ matrix related the magnetic 
impurity at
the left edge of the system is given by 
\begin{equation}
\exp \left[ i\varphi _a(k)\right] =\frac{2J_a(q+iC_a)\exp
(ik)+i[4+(4V_a-J_a)\exp (ik)]}{2J_a(q-iC_a)\exp (-ik)-i[4+(4V_a-
J_a)\exp
(-ik)]}.
\end{equation}
The phase factor $\exp \left[ i\varphi _b(k)\right] $, corresponding 
to the
scattering of the electron with the impurity at the right edge of the
system, is given by the substituting $b$ for $a.$ Without loss of
generality, the interaction parameters \ $J_{a,b}$ and $V_{a,b}$ take 
the
following forms: 
\begin{equation}
J_a=\frac 2{\left( s_a+c_a\right) \left( s_a-c_a+1\right) },  
\label{b001}
\end{equation}
\begin{equation}
V_a=-\frac{4c_a^2-\left( 2s_a+1\right) ^2-3}{4\left( s_a+c_a\right) 
\left(
s_a-c_a+1\right) }  \label{b002}
\end{equation}
The parameters $J_b$ and $V_b$ are given similarly by the substitution 
of
index $a$ by index $b$ in the above relations. By the use of these
parameterizations of the interaction constants $J_{a,b}$ and $V_{a,b}$ , 
the
phase factors take the form: $\exp \left[ i\varphi _a\left( k\right) 
\right]
=\exp \left[ i\varphi _b\left( k\right) \right] =-\exp \left( ik\right) 
$.
Then, the boundary $R$ matrices for the impurity model with the 
arbitrary
spins can be expressed as 
\begin{equation}
R_a(k_j,\sigma _j)=-\exp \left( ik_j\right) \frac{2q_j-2ic_a-i\left(
2s_a+1\right) P_{aj}}{2q_j+2ic_a+i\left( 2s_a+1\right) P_{aj}},  
\label{r003}
\end{equation}
\begin{equation}
R_b(-k_j,\sigma _j)=-\exp \left[ -ik_j\left( 2G+1\right) \right] 
\frac{
2q_j-2ic_b-i\left( 2s_b+1\right) 
P_{bj}}{2q_j+2ic_b+i\left( 2s_b+1\right)
P_{bj}}  \label{r004}
\end{equation}
with $q_j=\frac 12\tan \frac{k_j}2$, respectively. Here the operators 
$
P_{aj} $ and $P_{bj}$ are the generalized ``permutation operators'' and 
have
the expressions (\ref{p001}) and (\ref{p002}).

\section{Bethe Ansatz Equations of the Impurity Model}

Firstly, we can write down the wave function $\Psi _{\sigma _1,\sigma
_2,\cdots ,\sigma _N}(x_1,x_2,\cdots ,x_N)$ in the region $0\leq 
x_{Q1}\leq
x_{Q2}\leq \cdots \leq x_{QN}\leq G-1$ as 
\begin{eqnarray}
&&\Psi _{\sigma _1,\sigma _2,\cdots ,\sigma _N}(x_1,x_2,\cdots ,x_N) 
\nonumber \\
&=&\sum_P\sum_{r_1,r_2,\cdots r_N=\pm 1}\varepsilon _P\varepsilon
_rA_{\sigma _{Q1},\sigma _{Q2},\cdots ,\sigma
_{QN}}(r_{PQ1}k_{PQ1},r_{PQ2}k_{PQ2},\cdots ,r_{PQN}k_{PQN})  
\nonumber \\
&&\cdot \exp [i\sum_{j=1}^Nr_{Pj}k_{Pj}x_j],
\end{eqnarray}
where the coefficients $A_{\sigma _{Q1},\sigma _{Q2},\cdots ,\sigma
_{QN}}(r_{PQ1}k_{PQ1},r_{PQ2}k_{PQ2},\cdots ,r_{PQN}k_{PQN})$ are 
also
dependent on the spins of magnetic impurities which are suppressed for
brevity,and $\varepsilon _P=1(-1),$when the parity of $P$ is even(odd)$
,\varepsilon _r=\prod_{j=1}^Nr$ in which $r$ takes the value$+1$or 
$-1.$ The
coefficients $A$ are related by the scattering matrices. The boundary 
$R$
matrix satisfies the reflecting Yang-Baxter equation (\ref{byb} ). The
scattering matrices in the bulk satisfy the Yang-Baxter equation 
(\ref{facto}
). By the use of the standard Bethe ansatz method, we can diagonalize 
the
Hamiltonian with the two magnetic impurities\cite{last01,re04}. The 
boundary 
$R$ matrices can be cast into the forms: 
\begin{equation}
R_a(k_j,\sigma _j)=-\exp (ik_j)\frac{q_j-
ic_a+i\left( s_a+1/2\right) }{
q_j+ic_a-i\left( s_a+1/2\right) }\frac{S_{j0}(k_j,k_0)}{S_{j0}(-
k_j,k_0)},
\end{equation}
\begin{equation}
R_b(-k_j,\sigma _j)=-\exp \left[ -ik_j\left( 2G+1\right) \right] 
\frac{
q_j-ic_b+i\left( s_b+1/2\right) }{q_j+ic_b-
i\left( s_b+1/2\right) }\frac{
S_{jN+1}(k_j,k_{N+1})}{S_{jN+1}(-k_j,k_{N+1})},
\end{equation}
where 
\begin{equation}
S_{j0}(k_j,k_0)=-\frac{q_j-q_0-i\left( s_a+1/2\right) P_{aj}}{
q_j-q_0+i\left( s_a+1/2\right) },
\end{equation}
\begin{equation}
S_{jN+1}(k_j,k_{N+1})=-\frac{q_j-q_{N+1}-i\left( s_b+1/2\right) 
P_{bj}}{
q_j-q_{N+1}+i\left( s_b+1/2\right) }
\end{equation}
with $q_0=ic_a$ and $q_{N+1}=ic_b$. Notice that the scattering matrix 
in the
bulk has the form 
\begin{equation}
S_{jl}(k_j,k_l)=-\frac{q_j-q_l-iP_{jl}}{q_j-q_l+i},\ 
(j,l=1,2,\cdots ,N)
\end{equation}
where $q_j=\frac 12\tan \frac{k_j}2$. The present case ( $J=2$ and 
$V=3/2$ )
corresponds to triplet scattering and not to the ordinary $t-J$ model. 
The
symmetry of this model is $SU(3)$, rather than a graded $FFB$ 
super-algebra
which corresponds to the traditional $t-J$ model. In this 
case\cite{last01}, 
$\chi ^t=V/2+J/8=1$ and $\chi ^s=V/2-3J/8=0$. These quantities are
interchanged in Ref. \cite{last01}, but this was corrected in Ref.\cite
{review}. If we set 
\[
S_{\tau l}\left( \lambda \right) =-\frac{\lambda -q_l-iP_{\tau 
l}}{\lambda
-q_l+i},\ (l=1,2,\cdots ,N) 
\]
\begin{eqnarray}
S_{\tau 0}\left( \lambda \right) &=&-\frac{2\lambda -2ic_a-i\left(
2s_a+1\right) P_{a\tau }}{2\lambda -2ic_a+i\left( 2s_a+1\right) }, \\
S_{\tau N+1}\left( \lambda \right) &=&-\frac{2\lambda -2ic_b-i\left(
2s_b+1\right) P_{b\tau }}{2\lambda -2ic_b+i\left( 2s_b+1\right) },  
\nonumber
\end{eqnarray}
then we get the equation 
\begin{eqnarray}
\left. Tr\left[ T\left( \lambda \right) T^{-1}\left( -\lambda \right)
\right] \right| _{\lambda =q_j}\Phi &=&\exp \left( 2ik_jG\right) 
\frac{i-q_j
}{i/2+q_j}\frac{q_j+ic_a-i\left( s_a+1/2\right) }{q_j-ic_a+i\left(
s_a+1/2\right) }  \nonumber \\
&&\cdot \frac{q_j+ic_b-i\left( s_b+1/2\right) }{q_j-ic_b+i\left(
s_b+1/2\right) }\Phi  \label{ss003}
\end{eqnarray}
where $\Phi $ is the eigenstate of the system and $T(\lambda )$ is 
defined
as 
\begin{equation}
T\left( \lambda \right) =S_{\tau j}\left( \lambda \right) S_{\tau 
0}\left(
\lambda \right) S_{\tau 1}\left( \lambda \right) \cdots S_{\tau j-
1}\left(
\lambda \right) S_{\tau j+1}\left( \lambda \right) \cdots S_{\tau 
N+1}\left(
\lambda \right) .
\end{equation}
Then we have the following Bethe ansatz equations: 
\begin{eqnarray}
&&e\left( 2q_j\right) ^{2G}\prod_{\beta =1}^Me\left( 2q_j-2\lambda 
_\beta
\right) e\left( 2q_j+2\lambda _\beta \right)  \nonumber \\
&=&e\left( \frac{2q_j}{2c_a+2s_a+1}\right) 
e\left( \frac{2q_j}{2c_b+2s_b+1}
\right)  \nonumber \\
&&\cdot \prod_{l=1\left( l\neq j\right) }^Ne\left( q_j-q_l\right) 
e\left(
q_j+q_l\right) ,  \label{b004}
\end{eqnarray}
\begin{eqnarray}
&&e\left( \frac{\lambda _\alpha }{s_a+c_a}\right) 
e\left( \frac{\lambda
_\alpha }{s_b+c_b}\right) e\left( \frac{\lambda _\alpha }{s_a-
c_a}\right)
e\left( \frac{\lambda _\alpha }{s_b-c_b}\right)  \nonumber \\
&&\cdot \prod_{l=1}^Ne\left( 2\lambda _\alpha -2q_l\right) 
e\left( 2\lambda
_\alpha +2q_l\right)  \nonumber \\
&=&\prod_{\beta =1\left( \beta \neq \alpha \right) }^Me\left( \lambda
_\alpha -\lambda _\beta \right) e\left( \lambda _\alpha +\lambda _\beta
\right) ,  \label{m001}
\end{eqnarray}
where we have used the notation $e(x)=\left( x+i\right) /\left( x-
i\right) $ 
\cite{40100}. Notice that $e(\pm \infty )=1.$ For the case of $J=-2$ 
and $
V=-3/2,$ only the energy changes sign. The Bethe ansatz equations for 
this
case are the same as for antiferromagnetic coupling.

\section{Ground State Properties}

The eigenvalue of the Hamiltonian can be written down as 
\begin{equation}
E=2N-\sum_{j=1}^N\frac 1{q_j^2+\frac 14}
\end{equation}
with $q_j=\frac 12\tan \frac{k_j}2$. In order to obtain the ground state
properties for the case of the triplet state scattering we should find 
out
the solutions of the above Bethe ansatz equations. The sets of the
rapidities $\left\{ q_j\right\} $ and $\left\{ \lambda _\alpha \right\} 
$
have real and the complex solutions. Complex solutions of the $\lambda
_\alpha $ correspond to the excited states and will be discussed in 
section
6.

For $J=2$ and $V=\frac 32$, the model is the $SU(3)$ invariant $t-J$ 
model
where electrons in triplet states are scattered but not those in singlet
states. For the bulk this model is isomorphic to the spin $1$ Heisenberg
chain with $SU(3)$ invariance. It has no graded super-algebra and
corresponds to three bosonic degrees of freedom. All rapidities are real 
for
the ground state. Now we take the thermodynamic limit and introduce the
distribution functions $\rho \left( q\right) $ and $\sigma 
\left( \lambda
\right) $ corresponding to the parameters $\ q$ and $\lambda .$ The
corresponding ones of the holes are denoted by $\rho ^h\left( q\right) 
$ and 
$\sigma ^h\left( \lambda \right) $. So the integral equations 
describing the
ground state of the impurity model can be written as 
\begin{eqnarray}
&&a\left( q,\frac 12\right) +\frac 12\int d\lambda \sigma 
\left( \lambda
\right) \left[ a\left( q-\lambda ,\frac 12\right) 
+a\left( q+\lambda ,\frac 1
2\right) \right]   \nonumber \\
&=&\rho \left( q\right) +\rho ^h\left( q\right) +\frac 
1{2G}\left[ a\left(
q,c_a+s_a+\frac 12\right) +a\left( q,c_b+s_b+\frac 12\right) -
a\left( q,
\frac 12\right) \right]   \nonumber \\
&&+\frac 12\int \rho \left( q^{\prime }\right) 
dq^{\prime }\left[ a\left(
q-q^{\prime },1\right) +a\left( q+q^{\prime },1\right) \right] ,
\label{ss005}
\end{eqnarray}
\begin{eqnarray}
&&\frac 1{2G}\left[ a\left( \lambda ,s_a+c_a\right) +a\left( \lambda
,s_b+c_b\right) +a\left( \lambda ,s_a-c_a\right) +a\left( \lambda
,s_b-c_b\right) +a\left( \lambda ,\frac 12\right) \right]   \nonumber 
\\
&&+\frac 12\int dq\rho \left( q\right) \left[ a\left( \lambda -q,\frac 
12
\right) +a\left( \lambda +q,\frac 12\right) \right]   \nonumber \\
&=&\sigma \left( \lambda \right) +\sigma ^h\left( \lambda \right) 
+\frac 12
\int d\lambda ^{\prime }\sigma \left( \lambda ^{\prime }\right) \left[
a\left( \lambda -\lambda ^{\prime },1\right) +a\left( \lambda +\lambda
^{\prime },1\right) \right] ,  \label{ss006}
\end{eqnarray}
where $a\left( \lambda ,\eta \right) \equiv \frac 1\pi \frac \eta 
{\lambda
^2+\eta ^2}.$ The ground state properties can be obtained from the above
coupled integral equations. The details are as follows.

By taking into account of the above distribution functions, the energy 
per
site of the system can be expressed as 
\begin{equation}
\frac EG=\frac{2N}G-2\pi \int dq\rho \left( q\right) a\left( q,\frac 
12
\right)
\end{equation}
for the case of $J=2$ and $V=3/2.$ The number of the particles per site 
is 
\begin{equation}
\frac NG=\int dq\rho \left( q\right) .
\end{equation}
The magnetization of the impurity model can be written as 
\begin{equation}
\frac{S_z}G=\frac 12\int dq\rho \left( q\right) -\int d\lambda \sigma 
\left(
\lambda \right) +\frac{l_a+l_b}G
\end{equation}
where $l_a=1-s_a,$ $2-s_a,$ $\cdots ,$ $s_a$ and $l_b=1-s_b,$ $2-s_b,$ 
$
\cdots ,$ $s_b.$ $s_a$ and $s_b$ are the spins of the impurities located 
at
the two ends of the one-dimensional lattice system. The Fourier
transformations of the Bethe ansatz equations (\ref{ss005}) and 
(\ref{ss006}
) give that 
\begin{equation}
e^{-\left| \omega \right| /2}+e^{-\left| \omega \right| 
/2}\widetilde{\sigma 
}\left( \omega \right) =\widetilde{\rho }\left( \omega \right) \left(
1+e^{-\left| \omega \right| }\right) +\widetilde{\rho }^h\left( \omega
\right) +\frac 1{2G}\widetilde{\rho }_G\left( \omega \right) ,
\end{equation}
\begin{equation}
\widetilde{\rho }\left( \omega \right) e^{-\left| \omega \right| 
/2}+\frac 1{
2G}\widetilde{\sigma }_G\left( \omega \right) 
=\widetilde{\sigma }\left(
\omega \right) \left( 1+e^{-\left| \omega \right| }\right) 
+\widetilde{
\sigma }^h\left( \omega \right) ,
\end{equation}
where the tilde denotes the Fourier transform and the functions 
$\widetilde{
\rho }_G\left( \omega \right) $ and $\widetilde{\sigma }_G\left( \omega
\right) $ are the Fourier transformations of $\rho _G\left( q\right) 
$ and $
\sigma _G\left( \lambda \right) $ given by 
\begin{equation}
\rho _G\left( q\right) =a\left( q,c_a+s_a+\frac 12\right) 
+a\left( q,c_b+s_b+
\frac 12\right) -a\left( q,\frac 12\right) ,
\end{equation}
\begin{eqnarray}
\sigma _G\left( \lambda \right) &=&a\left( \lambda ,s_a+c_a\right) 
+a\left(
\lambda ,s_b+c_b\right) +a\left( \lambda ,s_a-c_a\right)  \nonumber 
\\
&&+a\left( \lambda ,s_b-c_b\right) +a\left( \lambda ,\frac 12\right) ,
\end{eqnarray}
which are the finite size corrections due to the magnetic impurities 
and the
open boundary conditions. Now we first consider the half-filled band 
of the
ferromagnetic case of $\sigma \left( \lambda \right) =0$. The nonzero
distribution functions are 
\begin{equation}
\widetilde{\rho }\left( \omega \right) =\frac 1{2\cosh \frac \omega 
2}-\frac 
1{2G}\frac{\widetilde{\rho }_G\left( \omega \right) }{1+e^{-\left| 
\omega
\right| }},
\end{equation}
\begin{equation}
\widetilde{\sigma }^h\left( \omega \right) =\frac{e^{-\left| \omega 
\right|
/2}}{2\cosh \frac \omega 2}+\frac 
1{2G}\left( \widetilde{\sigma }_G\left(
\omega \right) -\frac{\widetilde{\rho }_G\left( \omega 
\right) }{2\cosh 
\frac \omega 2}\right)
\end{equation}
and the energy is 
\begin{equation}
E/G=1-2\ln 2.
\end{equation}
The finite size correction to the energy in this case is 
\begin{equation}
E_{fin}^0=\ln 2+\frac 12\int_0^\infty 
\frac{\widetilde{\rho }_G\left( \omega
\right) }{\cosh \frac \omega 2}d\omega ,
\end{equation}
contributed by the impurities, and the open boundary term gives the
correction $-\ln 2$. For the nonmagnetic case, when the number per site 
is $
2/3,$ the distribution functions take the form: 
\begin{equation}
\widetilde{\rho }\left( \omega \right) =\frac{2\cosh \frac \omega 2}{
1+2\cosh \omega }+\frac 1{2G}\frac{e^{\left| \omega \right| 
/2}}{1+2\cosh
\omega }\left\{ \widetilde{\sigma }_G\left( \omega \right) -
2\widetilde{\rho 
}_G\left( \omega \right) \cosh \frac \omega 2\right\} ,
\end{equation}
\begin{equation}
\widetilde{\sigma }\left( \omega \right) =\frac 1{1+2\cosh 
\omega }+\frac 1{
2G}\frac{e^{\left| \omega \right| /2}}{1+2\cosh 
\omega }\left\{ 2\widetilde{
\sigma }_G\left( \omega \right) \cosh \frac \omega 2-\widetilde{\rho }
_G\left( \omega \right) \right\}
\end{equation}
with the energy being 
\begin{equation}
\frac EG=\frac 43-\frac{\pi \sqrt{3}}9-\ln 3,
\end{equation}
and the finite size correction of the energy is 
\begin{equation}
E_{fin}^1=\frac{2\pi \sqrt{3}}9-\frac 12\int_{-\infty }^\infty \frac{
\widetilde{\sigma }_G\left( \omega \right) -
2\widetilde{\rho }_G\left(
\omega \right) \cosh \frac \omega 2}{1+2\cosh \omega }d\omega .
\end{equation}
This is the contribution of the magnetic impurities and the open 
boundary
terms give the correction $-2\pi \sqrt{3}/9$.

\section{Excited State}

The eigenstates of the model are specified by two sets of rapidities, 
$
\{q_j\}$ and $\{\lambda _j\},$ for the charges and the spins, 
respectively.
In this section, we will also restrict the discussions on the case of 
$J=2$
and $V=3/2.$ In the thermodynamic limit these rapidities are classified
according to the string hypothesis:\cite{schlo02,2525} 
\[
q_\alpha ^{n,j}=q_\alpha ^n+\frac i2\left( n+1-2j\right) ,\ 
j=1,2,\cdots ,n
\]
\begin{equation}
\lambda _\beta ^{n,j}=\lambda _\beta ^n+\frac i2\left( n+1-
2j\right) ,\
j=1,2,\cdots ,n  \label{ss0106}
\end{equation}
where $n=1,2,\cdots ,+\infty .$ These strings describe charge and spin
boundstates, respectively. The real parameters $q_\alpha ^n$ and 
$\lambda
_\beta ^n$ are related to the momentum of the center of mass of the
boundstate. Their distribution functions are $\rho _n\left( q\right) 
$, $
\sigma _n\left( \lambda \right) $ and the ones corresponding to the 
holes
are $\rho _n^h\left( q\right) $ and $\sigma _n^h\left( \lambda 
\right) .$ By
using the notations

\begin{equation}
\left[ n\right] f(k)=\int_{-\infty }^\infty a\left( k-
k^{\prime },\frac n2
\right) f(k^{\prime })dk^{\prime }\text{\qquad for }n\neq 0,  
\label{ddd1}
\end{equation}
\[
\left[ 0\right] f(k)\equiv f(k), 
\]
the integral equations can be cast into the following compact forms: 
\[
a\left( q,\frac n2\right) +\sum_{m=1}^\infty \sum_{j=0}^{m-1}\left[
n-m+1+2j\right] \sigma _m\left( q\right) +\frac 1{2G}\rho _n^G\left(
q\right) 
\]
\begin{equation}
\qquad \qquad \qquad \qquad \qquad =\rho _n^h\left( q\right)
+\sum_{m=1}^\infty A_{nm}\rho _m\left( q\right) ,  \label{ss013}
\end{equation}
\begin{equation}
\frac 1{2G}\sigma _n^G\left( \lambda \right) +\sum_{m=1}^\infty
\sum_{j=0}^{m-1}\left[ n-m+1+2j\right] \rho _m\left( \lambda \right) 
=\sigma
_n^h\left( \lambda \right) +\sum_{m=1}^\infty A_{nm}\sigma 
_m\left( \lambda
\right)  \label{ss014}
\end{equation}
where the terms with the factor $1/\left( 2G\right) $ are the 
contributions
provided by the magnetic impurities and the boundary terms. The operator 
$
A_{nm}$ is defined as 
\begin{equation}
A_{nm}\equiv \left[ \left| n-m\right| \right] +2\left[ \left| n-
m\right|
+2\right] +2\left[ \left| n-m\right| +4\right] +\cdots 
+2\left[ n+m-2\right]
+\left[ n+m\right]
\end{equation}
and 
\begin{equation}
\rho _n^G\left( q\right) \equiv a\left( q,\frac n2\right)
-\sum_{j=0}^{n-1}\left\{ a\left( q,s_a+c_a-j+\frac n2\right) +a\left(
q,s_b+c_b-j+\frac n2\right) \right\} ,  \label{ss017}
\end{equation}
\begin{eqnarray}
\sigma _n^G\left( \lambda \right) &=&a\left( \lambda ,\frac n2\right)
+\sum_{j=0}^{n-1}\left\{ a\left( \lambda ,s_a+c_a-j+\frac{n-
1}2\right)
+a\left( \lambda ,s_b+c_b-j+\frac{n-1}2\right) \right.  \nonumber \\
&&\left. +a\left( \lambda ,s_a-c_a-j+\frac{n-1}2\right) 
+a\left( \lambda
,s_b-c_b-j+\frac{n-1}2\right) \right\}  \label{ss018}
\end{eqnarray}
Now the number of the electrons per site is expressed by 
\begin{equation}
\frac NG=\sum_{m=1}^\infty m\int \rho _m\left( q\right) dq.
\end{equation}
By minimizing the thermodynamic potential we get the thermodynamical 
Bethe
ansatz equations: 
\[
\frac{n\left( 2-A-H\right) -2\pi a\left( k,\frac n2\right) }T-\ln 
\left(
1+\xi _n\right) +\sum_{m=1}^\infty A_{nm}\ln \left( 1+\xi _m^{-
1}\right) 
\]
\begin{equation}
\qquad \qquad \qquad -\sum_{m=1}^\infty \sum_{j=0}^{n-1}\left[
m-n+1+2j\right] \ln \left( 1+\eta _m^{-1}\right) =0,  \label{t001}
\end{equation}
\[
\frac{2Hn}T-\ln \left( 1+\eta _n\right) +\sum_{m=1}^\infty A_{nm}\ln 
\left(
1+\eta _m^{-1}\right) \qquad \qquad 
\]
\begin{equation}
\qquad \qquad -\sum_{m=1}^\infty \sum_{j=0}^{n-1}\left[ m-
n+1+2j\right] \ln
\left( 1+\xi _m^{-1}\right) =0,  \label{t002}
\end{equation}
where we have set that 
\[
\eta _n\left( k\right) =\frac{\sigma _n^h\left( k\right) }{\sigma 
_n\left(
k\right) },\quad \xi _n\left( k\right) =\frac{\rho 
_n^h\left( k\right) }{
\rho _n\left( k\right) }. 
\]
Of course, the above expressions can be changed into the several other 
forms
as the ordinary system in one dimensional lattice system. The impurities
located at the ends of the system do not change the free energy for the 
bulk
\cite{schlo02,2525}. The finite size correction of the free energy
contributed by the boundary term is 
\begin{equation}
F_{bou}=-\frac T2\sum_{n=1}^\infty \left[ n\right] \ln \left\{ 1+\xi
_n^{-1}\left( 0\right) \right\} -\frac T2\sum_{n=1}^\infty 
\left[ n\right]
\ln \left\{ 1+\eta _n^{-1}\left( 0\right) \right\} .
\end{equation}
The free energy due to the two magnetic impurities with the spins 
$s_{a,b}$
is 
\begin{eqnarray}
F_{imp} &=&-\frac T2\sum_{n=1}^\infty \int dq\rho _{n,imp}^G\ln 
\left( 1+\xi
_n^{-1}\right) -2H\left( l_a+l_b\right)  \nonumber \\
&&-\frac T2\sum_{n=1}^\infty \int d\lambda \sigma _{n,imp}^G\ln \left(
1+\eta _n^{-1}\right) ,
\end{eqnarray}
where $l_a=1-s_a,$ $2-s_a,$ $\cdots ,$ $s_a$; $l_b=1-s_b,$ $2-s_b,$ 
$\cdots
, $ $s_b;$ and 
\begin{equation}
\rho _{n,imp}^G(q)=-\sum_{j=0}^{n-1}\left\{ a\left( q,s_a+c_a-
j+\frac n2
\right) +a\left( q,s_b+c_b-j+\frac n2\right) \right\} ,
\end{equation}
\begin{equation}
\sigma _{n,imp}^G(\lambda )=\sum_{r=\pm 1,l=a,b}\sum_{j=0}^{n-
1}a\left(
\lambda ,s_l+rc_l-j+\frac{n-1}2\right) .
\end{equation}
\quad The interesting observation that the finite size correction due 
to the
magnetic impurities is finite even if the temperature $T\rightarrow 0.$ 
The
properties of the system at the high-temperature limit and the low
temperature limit can be discussed by the use of the thermal Bethe ansatz
equations. And the boundary bound states exist in our model. These will 
be
studied in the following section.

\section{Special Limits and the Boundary Bound States}

In order to discuss the low temperature properties of the model and the
impurity free energy at high-temperature, we may change the thermal 
Bethe
ansatz equations into the forms: 
\begin{equation}
\frac{2\pi \left( \left[ 0\right] +\left[ 2\right] \right) ^{-
1}a(k,\frac 12)
}T+\widehat{G}\ln \frac{1+\eta _2^{-1}}{1+\xi _2}+\ln \xi _1=0,
\end{equation}
\begin{equation}
\ln \xi _n=\widehat{G}\ln \frac{\left( 1+\xi _{n-1}\right) 
\left( 1+\xi
_{n+1}\right) }{(1+\eta _{n-1}^{-1})(1+\eta _{n+1}^{-1})},\quad n>1,
\end{equation}
\begin{equation}
\frac{2\pi \left( \left[ 0\right] +\left[ 2\right] \right) ^{-
1}a(k,\frac 12)
}T-\widehat{G}\ln \left( \xi _2\eta _2\right) +\ln \left( \xi _1\eta
_1\right) =0,
\end{equation}
\begin{equation}
\ln \eta _n=\widehat{G}\ln \frac{(1+\eta _{n-1})(1+\eta 
_{n+1})}{\left(
1+\xi _{n-1}^{-1}\right) \left( 1+\xi _{n+1}^{-1}\right) },\quad n>1,
\end{equation}
\begin{equation}
\lim_{n\rightarrow \infty }\left( \left[ n+1\right] \ln \frac{1+\xi 
_n}{
1+\eta _n^{-1}}-\left[ n\right] \ln \frac{1+\xi _{n+1}}{1+\eta 
_{n+1}^{-1}}
\right) =\frac{A-2+H/2}T
\end{equation}
\begin{equation}
\lim_{n\rightarrow \infty }\left( \left[ n+1\right] \ln \frac{1+\eta 
_n}{
1+\xi _n^{-1}}-\left[ n\right] \ln \frac{1+\eta _{n+1}}{1+\xi 
_{n+1}^{-1}}
\right) =-\frac HT
\end{equation}
with the use of the formulae $A_{1m}-\widehat{G}A_{2m}=\delta _{1m},$ 
$
A_{nm}-\widehat{G}\left( A_{n-1\ m}+A_{n+1\ m}\right) =\delta _{nm}$ 
$(n>1),$
and $\left[ n+1\right] A_{nm}-\left[ n\right] A_{n+1\ m}=0$ if $m<n$ 
or $
-\left[ m+1\right] -\left[ m-1\right] $ if $m>n$ where $\widehat{G}
=[1]/([0]+[2]).$ Furthermore, we have that $\lim_{n\rightarrow \infty
}\left( \ln \xi _n\right) /n=(2-A-H/2)/T$ and $\lim_{n\rightarrow 
\infty
}\left( \ln \eta _n\right) /n=H/T$ where $H$ is the magnetic field and 
$A$
is the chemical potential of the system. Define the symbol function 
$sign(a)$
as 
\[
sign(a)=\left\{ 
\begin{array}{c}
1\quad a>0 \\ 
0\quad a=0 \\ 
-1\quad a<0
\end{array}
\right. . 
\]
Then, in the high-temperature limit, the free energy of the impurities 
is 
\begin{eqnarray}
F_{imp} &=&\frac T2\ln 2\sum_{l=a,b}\sum_{r_1=0,\pm 
1}\left\{ sign\left(
s_l+c_l+\frac 12+r_1\right) \right.  \nonumber \\
&&\left. -\sum_{r_2=\pm 1}sign\left( s_l+r_2c_l+r_1\right) \right\} .
\end{eqnarray}
The boundary term give the contribution $F_{bou}=-3T\ln 2.$ The above 
result
means that the entropy of the model related to the impurities is 
dependent
on the interacting parameters of the impurities with the electrons and 
the
spins of the impurities. At the low temperature limit, from the thermal
Bethe ansatz equations, we get that \cite{yulu,tsve} the functions $\xi 
_n$
have the different asymptotic properties for $n<2$ and $n\geq 2$. It 
means
that \cite{yulu} $f=2$ and this gives that $\tau \equiv 
4/\left( f+2\right)
=1.$ By considering the free energy expression in the bulk, we find that 
at
low temperature, the specific heat behaves as $C_{bul}\sim T.$ Since 
the
asymptotic property of the term $\ln (1+\eta _n^{-1})$ are different 
from
the one of the term $\ln (1+\xi _n^{-1})$ at the low temperature limit, 
the
impurities and the open boundary term contributions to the specific heat 
at
low temperature are nonlinear at the temperature. Although it is
difficult to get the exact expression of the specific heat, we know that
this nonlinear relation with the temperature is related to the 
interacting
parameters and the spins of the impurities.

The boundary bound states can be formed in the present model in the 
charge
sectors. When $\left| s_a+c_a\right| <\frac 12,$ the boundary bound 
state
can be denoted by 
\begin{eqnarray}
q^{3,0} &=&\pm i\left( s_a+c_a+\frac 12\right) ,  \nonumber \\
q^{3,1} &=&\mp \frac i2\left( s_a+c_a-\frac 12\right) , \\
q^{3,2} &=&\mp \frac i2\left( s_a+c_a+\frac 32\right)  \nonumber
\end{eqnarray}
in the charge sectors. It carries the energy: 
\begin{equation}
E_{edg}^{(3)}=\frac{3(2s_a+2c_a+1)^2(2s_a^2+2c_a^2+4s_ac_a+2s_a+2c_
a-5)}{
(s_a+c_a)(s_a+c_a+1)(2s_a+2c_a-3)(2s_a+2c_a+5)}.
\end{equation}
Obviously, when we substitute the subindex $a$ by $b$, we also get the 
bound
state with three imaginary modes localized at the end of the system. 
The
total momentum of the bound states are zero. Another kind of the boundary
bound state with five imaginary modes can be expressed as 
\begin{eqnarray}
q^{5,0} &=&\pm i\left( s_a+c_a+\frac 12\right) ,  \nonumber \\
q^{5,1} &=&\mp \frac i2\left( 3s_a+3c_a+\frac 12\right) ,  \nonumber 
\\
q^{5,2} &=&\mp \frac i2\left( 3s_a+3c_a+\frac 52\right) , \\
q^{5,3} &=&\pm i\left( s_a+c_a+\frac 32\right) ,  \nonumber \\
q^{5,4} &=&\pm i\left( s_a+c_a-\frac 12\right)  \nonumber
\end{eqnarray}
when $-\frac 12<s_a+c_a<-\frac 16.$ It has the energy 
\begin{equation}
E_{edg}^{(5)}=\frac{5\left(
72(s_a+c_a)^4+144(s_a+c_a)^3-58(s_a+c_a)^2-
130(s_a+c_a)+11\right) }{
(s_a+c_a-1)(s_a+c_a+2)(6s_a+6c_a-1)(6s_a+6c_a+7)}.
\end{equation}
The imaginary modes with the subindex $b$ also form the boundary bound
state. The momentum of the state, which is the sum of the ones of the
imaginary modes, is also zero. These bound states are at the ends of 
the
system.

\section{Concluding Remarks}

Recently, in an interesting work\cite{prlzv}, Zvyagin get that the
low-energy magnetic behaviors of an impurity in a chain with periodic
boundary conditions and with open boundary conditions coincide up to
mesoscopic corrections of order of $G^{-1}$ and it was shown that this
property is independent of the impurity position in the open chain. From 
the
above discussions, we get the same conclusion. The impurity part of the
Hamiltonian for the open boundary condition in Ref. \cite{prlzv}is 
\begin{equation}
H_{imp}=[K_{\alpha \beta }/(1+u_0^2)](J_0^\alpha J_1^\beta +H.c.)
\end{equation}
with the ``weak link'' $(1+u_0^2)^{-1}<1$ where $K_{\alpha \beta
}=StrJ^\alpha J^\beta $ and $J_{0,1}^{\alpha =1,\cdots ,9}$ are the
generators of the supersymmetric algebra $sl(1|2)$. In our model, for 
the
open boundary conditions, the impurity part of the Hamiltonian is 
\begin{equation}
H_{imp}=J_a{\bf S}_1\cdot {\bf d}_a+V_an_1+J_b{\bf S}_G\cdot {\bf d}
_b+V_bn_G.
\end{equation}
$J_{a,b}$ and $V_{a,b}$ depend on the interacting parameters $c_{a,b}$ 
and
the spins $s_{a,b}$ of the impurities (See relations (\ref{b001}-
\ref{b002}
)). The finite size corrections of the ground state due to the impurities
are obtained and the properties of the system at the high-temperature 
limit
and the low temperature limit are studied in this paper. We also 
discussed
the boundary bound states which are localized at the edges of the system.

To conclude, we have constructed from the open boundary $t-J$ model a
Hamiltonian with the two magnetic impurities which have arbitrary 
spins .
The impurities are coupled to the ends of the system. The interaction
parameters of the impurities with the electrons are parameterized by 
the
constants which related on the potentials of the impurities in order 
to
preserve the integrability of the system. The boundary $R$ matrices of 
the
impurity model are given explicitly and satisfy the reflection equation. 
In
fact, our impurity model has also the following $R$ matrices, 
\begin{equation}
R_a(k_j,\sigma _j)=-\exp \left[ ik_j\mp i\theta _a\left( \pi -
k_j\right)
\right] \frac{2q_j-2ic_a-i\left( 2s_a+1\right) 
P_{aj}}{2q_j+2ic_a+i\left(
2s_a+1\right) P_{aj}},  \label{r007}
\end{equation}
\begin{equation}
R_b(-k_j,\sigma _j)=-\exp \left[ -ik_j\left( 2G+1\right) \mp i\theta
_b\left( \pi -k_j\right) \right] \frac{2q_j-2ic_b-
i\left( 2s_b+1\right)
P_{bj}}{2q_j+2ic_b+i\left( 2s_b+1\right) P_{bj}}  \label{r008}
\end{equation}
with $q_j=\frac 12\tan \frac{k_j}2$. Here the interaction parameters 
are
denoted by 
\[
J_a=\frac 2{\left( s_a-c_a\right) \left( s_a+c_a+1\right) }, 
\]
\[
V_a=-\frac{4c_a^2-\left( 2s_a+1\right) ^2-3}{4\left( s_a-c_a\right) 
\left(
s_a+c_a+1\right) }. 
\]
The function $\theta $ is defined as 
\[
\theta _a\left( k\right) =\frac 1i\ln \frac{\left( 2c_a^2-2s_a^2-
2s_a\right)
\cos k+2s_a^2-2c_a^2+2s_a+1+2ic_a\sin k}{\left( 2c_a^2-2s_a^2-
2s_a\right)
\cos k+2s_a^2-2c_a^2+2s_a+1-2ic_a\sin k}. 
\]
The parameters $J_b,$ $V_b$ and the function $\theta _b\left( k\right) 
$ can
be obtained from the above relations with the substitutions of the index 
$b$
for $a$ . By comparing the expressions of $J_{a,b}$ and $V_{a,b}$ with 
the
ones in section 3, we know that the different expressions of the boundary 
$R$
matrices are because of the transformations $c_a\rightarrow -c_a$ and 
$
c_b\rightarrow -c_b.$ This reminds us the $R$ matrices of the impurity 
model
can be written down as the other forms with similar substitutions.

The Hamiltonian including the arbitrary spin impurities is diagonalized
exactly by the Bethe ansatz method. The coupled integral equations of 
the
ground state are derived and the ground state properties are discussed 
in
details. The distribution functions of the rapidities are given also.
Furthermore, we study the string solution of the Bethe ansatz equations. 
The
integral equations of the excited state are obtained. By minimizing the
thermodynamic potential, we get the thermodynamic Bethe ansatz 
equations.
The finite size correction due to the magnetic impurities is obtained 
for
the free energy. It is interesting that this correction is finite when 
the
temperature $T\rightarrow 0,$ but have the value related to the
magnetization of the impurities and the external magnetic field. We
discussed also the properties of the system at the high-temperature 
limit
and the low temperature limit. Two kinds of the boundary bound states 
are
constructed also for the case of $J=2$ and $V=3/2.$ Finally, we point 
out
the impurity contributions are derived in the above discussion under 
the
thermodynamic limits since the Bethe ansatz equations can be changed 
into
the coupled integral equations and there solved by a Fourier transform. 
So,
it is an interesting problem to discuss the impurity effects for the 
finite
lattice case. de Vega and Woynarovich\cite{Vega2} have given a method 
to
calculate the leading-order finite size corrections to the ground state
energy. By using the finite size scaling technique 
\cite{10301,10302,10303},
the case of some other integrable models such as Hubbard model and
Heisenberg spin chain are also studied\cite{ham01,ham02,2996}. It is 
worth
studying the impurity effects in different sectors of the present model
further and its critical properties are also interesting topics for 
further
discussions.

\section*{Appendix A. The Boundary $R$ Matrices}

When the interacting parameters $J_{a,b}$ and $V_{a,b}$ take the forms 
\begin{equation}
J_a=\frac 2{\left( s_a-c_a\right) \left( s_a+c_a+1\right) },  
\eqnum{A.1}
\end{equation}
\begin{equation}
V_a=\frac{4c_a^2-\left( 2s_a+1\right) ^2+1}{4\left( s_a-c_a\right) 
\left(
s_a+c_a+1\right) }  \eqnum{A.2}
\end{equation}
for $J=2,$ $V=-1/2,$ the boundary $R$ matrices for the impurity model 
with
the arbitrary spins can be written down as

\begin{equation}
R_a(k_j,\sigma _j)=\exp \left( ik_j\right) \frac{2q_j-2ic_a-i\left(
2s_a+1\right) P_{aj}}{2q_j+2ic_a+i\left( 2s_a+1\right) P_{aj}},  
\eqnum{A.3}
\end{equation}
\begin{equation}
R_b(-k_j,\sigma _j)=\exp \left[ -ik_j\left( 2G+1\right) \right] \frac{
2q_j-2ic_b-i\left( 2s_b+1\right) 
P_{bj}}{2q_j+2ic_b+i\left( 2s_b+1\right)
P_{bj}}  \eqnum{A.4}
\end{equation}
with $q=\frac 12\cot \frac k2$ . When $J=-2$ and $V=1/2,$ $J_a$ and $V_a$
are 
\begin{equation}
J_a=-\frac 2{\left( s_a-c_a\right) \left( s_a+c_a+1\right) },  
\eqnum{A.5}
\end{equation}
\begin{equation}
V_a=-\frac{4c_a^2-\left( 2s_a+1\right) ^2+1}{4\left( s_a-c_a\right) 
\left(
s_a+c_a+1\right) }.  \eqnum{A.6}
\end{equation}
When $J=-2$ and $V=-3/2,$ we have that 
\begin{equation}
J_a=-\frac 2{\left( s_a+c_a\right) \left( s_a-c_a+1\right) },  
\eqnum{A.7}
\end{equation}
\begin{equation}
V_a=\frac{4c_a^2-\left( 2s_a+1\right) ^2-3}{4\left( s_a+c_a\right) 
\left(
s_a-c_a+1\right) }.  \eqnum{A.8}
\end{equation}
The boundary $R$ matrices are same as the relations (\ref{r003} -
\ref{r004}).

\section*{Appendix B. The Bethe Ansatz Equations}

From the standard Bethe ansatz procedure, we get the following Bethe 
ansatz
equations: 
\[
\exp \left( 2Gik_j\right) \frac{\cot 
\frac{k_j}2+2ic_a+i\left( 2s_a+1\right) 
}{\cot \frac{k_j}2-2ic_a-i\left( 2s_a+1\right) }\frac{\cot 
\frac{k_j}2
+2ic_b+i\left( 2s_b+1\right) }{\cot \frac{k_j}2-2ic_b-
i\left( 2s_b+1\right) }
\qquad 
\]
\begin{equation}
\qquad \qquad \qquad \qquad =\prod_{\beta =1}^M\frac{\cot \frac{k_j}2
-2\lambda _\beta +i}{\cot \frac{k_j}2-2\lambda _\beta -i}\frac{\cot 
\frac{
k_j }2+2\lambda _\beta +i}{\cot \frac{k_j}2+2\lambda _\beta -i},\qquad
(j=1,2,\cdots ,N)  \eqnum{B.1}
\end{equation}
\begin{eqnarray*}
&&\frac{\lambda _\alpha +ic_a+is_a}{\lambda _\alpha +ic_a-
is_a}\frac{\lambda
_\alpha +ic_b+is_b}{\lambda _\alpha +ic_b-is_b}\frac{\lambda _\alpha
-ic_a+is_a}{\lambda _\alpha -ic_a-is_a}\frac{\lambda _\alpha -
ic_b+is_b}{
\lambda _\alpha -ic_b-is_b} \\
&&\cdot \prod_{l=1}^N\frac{2\lambda _\alpha -\cot 
\frac{k_l}2+i}{2\lambda
_\alpha -\cot \frac{k_l}2-i}\frac{2\lambda _\alpha +\cot 
\frac{k_l}2+i}{
2\lambda _\alpha +\cot \frac{k_l}2-i}
\end{eqnarray*}

\begin{equation}
=\prod_{\beta =1(\beta \neq \alpha )}^M\frac{\lambda _\alpha -\lambda 
_\beta
+i}{\lambda _\alpha -\lambda _\beta -i}\frac{\lambda _\alpha +\lambda 
_\beta
+i}{\lambda _\alpha +\lambda _\beta -i}\qquad (\alpha =1,2,\cdots ,M) 
\eqnum{B.2}
\end{equation}
for the case of $J=2$ and $V=-1/2.$ When $J=-2$ and $V=1/2,$ the Bethe
ansatz equations are the same as the above ones, but the energy changes
sign. The Bethe ansatz equations for the case of $J=-2$ and $V=-3/2$ 
are the
same as (\ref{b004}-\ref{m001}) and the energy changes sign also.

\end{document}